\documentstyle[amsfonts,12pt,oldlfont]{article}

\title{Toda-Darboux maps and vertex operators}

\author{M. Adler\thanks{Department of Mathematics,
Brandeis University, Waltham, Mass 02254,
USA. The support of a National Science Foundation grant \#
DMS-9503246 is gratefully acknowledged.}~~~~~P. van
Moerbeke\thanks{Universit\'e de Louvain, 1348 Louvain-la-Neuve, Belgium
and Department of Mathematics, Brandeis University, Waltham, Mass
02254, USA. The  support of a National Science
Foundation grant \# DMS-9503246, a Nato, a FNRS
and a Francqui Foundation grant is gratefully
acknowledged.}}

\newcommand{\MAT}[1]{\left(\begin{array}{*#1c}}
\newcommand{\mat}{\end{array}\right)}

\newcommand{\qed}
{%
\mbox{}%
\nolinebreak%
\hfill%
\rule{2mm}{2mm}%
\medbreak%
\par%
}

\newcommand{\rg}{\rightarrow}

\newcommand{\DF}{\Longleftrightarrow}
\newcommand{\pp}{\ldots}
\newcommand{\DR}{{\cal D}}

\newcommand{\BZ}{{\Bbb Z}}
\newcommand{\BC}{{\Bbb C}}

\newcommand{\iy}{\infty}

\newcommand{\pl}{\partial}
\newcommand{\al}{\alpha}
\newcommand{\be}{\beta}
\newcommand{\proof}{\underline{\sl Proof}: }
\newcommand{\om}{\omega}

\newcommand{\ga}{\gamma}

\newcommand{\vr}{\varepsilon}

\newcommand{\lb}{\lambda}
\newcommand{\Lb}{\Lambda}
\newcommand{\BX}{{\Bbb X}}

\def\BY{{\Bbb Y}}

\def\be{\begin{equation}}
\def\ee{\end{equation}}
\def\bea{\begin{eqnarray}}
\def\eea{\end{eqnarray}}

\ifx\undefined\Bbb
        \let\Bbb\bf
\fi
\ifx\undefined\curvearrowright
        
\else
        
\fi

\catcode`\@=11
\def\ps@X{\let\@mkboth\@gobbletwo
        \def\@oddhead{\tt Adler-Van Moerbeke:%
        Toda-Darboux\hfil\today\ \#1\hfil\S\thesection, p.\thepage
        }
        \def\@oddfoot{\rm\hfil\thepage\hfil}
        \let\@evenhead\@oddhead
        \let\@evenfoot\@oddfoot}
\catcode`@=12
\begin{document}
\maketitle

\setcounter{equation}{0}

Consider the Toda lattice
\be
\frac{\pl L}{\pl t_n}=[( L^n)_+,
L]=[-( L^n)_-, L],\hspace{1cm}n=1,2,\pp\,\,.
\ee
on bi-infinite tridiagonal matrices
\be
L=\Lambda^{-1} a +b \Lambda^0 +\Lambda,
\ee
with $\Lambda$ being the customary shift $\Lambda
:=(\delta_{i,j-1})_{i,j\geq 0}$ and
$a$ and
$b$  being diagonal matrices. As is well known, in analogy with Sato's
KP-theory, the entries $a$ and $b$ have the following $\tau$-function
representation
\be
b_n=\frac{\pl}{\pl t_1} \log \frac {\tau_{n+1}}{\tau_n} ~\mbox{ and
}~ a_{n-1} = \frac {\tau_{n-1} \tau_{n+1}}{\tau_n^2},
\ee
in terms of a $\tau$-vector $\tau=(\tau_n)_{n\in\BZ}$. In section
2, it is shown that, if
$\tau$ is a Toda lattice $\tau$-vector, then
$$
(1+c\,\BX(t,z))\tau
$$
is a Toda $\tau$-vector as well, where the {\em vertex operator }
$X(t,z)$, first introduced by us in \cite{AvM2}, is given
by\footnote{$\chi(z):=\mbox{diag}~(...,z^{-1},1,z,...)$}
\be
\BX(t,z)=\Lb^{-1}\chi(z^2)e^{\sum t_iz^i}e^{-2\sum\frac{z^{-i}}{i}
\frac{\pl}{\pl t_i}}.
\ee
For a fixed $\lb\in\BC^*$, we consider a Borel factorization
$$L(t)-\lb=L_-(t)L_+(t),\mbox{\,\,with\,\,}L_-(t)=\Lambda^{-1}\al(t)
+\Lambda^0,\,\,L_+(t)=
\beta(t)\Lambda^0+\Lambda
$$
and the corresponding {\em Darboux transform}
\be
L(t)-\lb=L_-L_+\mapsto\tilde L(t)-\lb=L_+L_-,
\ee
which in coordinates reads:
\begin{eqnarray}
b_n-\lambda &=&\al_{n-1}+\beta_n\longmapsto\tilde
b_n-\lambda =\al_n+\beta_n=(b_n-\lambda)+ (\al_n-\al_{n-1})\nonumber\\
a_{n-1}&=&\al_{n-1}\beta_{n-1}\longmapsto\tilde
a_{n-1}=\al_{n-1}\beta_n=a_{n-1}\frac{\beta_n}{\beta_{n-1}}.
\end{eqnarray}
This map will be called {\it Toda-Darboux}, when both matrices
$L-\lb=L_-L_+$, and $\tilde L=L_+L_-$ flow according to all Toda vector
fields.

The following {\it vertex operators} will play an important role in
this paper:
\be
\BX_1(t,\lb):=\chi(\lb)X(t,\lb)\quad\mbox{and}\quad\BX_2(t,\lb)
:=\chi(\lb^{-1}) X(-t,\lb)\Lambda,
\ee
where
\be
X(t,\lb):=e^{\sum_1^{\iy}t_i\lb^i}e^{-\sum_1^{\iy}
\frac{\lb^{-i}}{i}\frac{\pl}{\pl t_i}}.
\ee
We also define a {\it discrete Wronskian} $\{\,,\,\}$ on column
vectors $f$ and $g$
\be
\{f,g\}:=(f_{n+1}g_n-f_ng_{n+1})_{n\in\BZ}.
\ee

\proclaim Theorem 0.1. Each element of the
2-dimensional null-space
\be
\ker(L(t)-\lb)=\left\{\Phi(t,\lb)=\frac{\tilde\tau(t)}{\tau(t)}=
\frac{\left(a\,\BX_1(t,\lb)+b\,e^{\sum
t_i\lb^i}\BX_2(t,\lb)\right)\tau(t)}{\tau(t)}
\right\},
\ee
where $\Phi(t,\lb)$ satisfies
$$
\frac{\pl\Phi}{\pl t_n}=(L^n)_+\Phi,
$$
specifies a factorization for all $t \in \BC^{\iy}$,
$$
L(t)-\lb=L_-L_+=(\Lb^{-1}\al+\Lb^0)(\beta\Lb^0+\Lb),
$$
with
\be
\al=\Lb\frac{\pl}{\pl
t_1}\log\Phi(t,\lb)-\lb\mbox{\,\,and\,\,}\beta=-
\frac{\Lb\Phi(t,\lb)}{\Phi(t,\lb)},
\ee
and so
$$
\ker L_+=\BC\Phi(t,\lb).
$$
The Darboux transform (0.5) maps $L(t)-\lb$ into a new tridiagonal
matrix $\tilde L(t)-\lb$, with entries $\tilde b_n$ and $\tilde
a_n$ given by (0.3) in term of
$\tilde\tau$. The Toda-Darboux transform on $L$ induces a map on $\tau$:
\be
\tau\longmapsto\tilde \tau=\tau\Phi=\left(a\BX_1(t,\lb)+b\,e^{\sum
t_i\lb^i}
\BX_2(t,\lb)\right)\tau(t).
\ee

\vspace{1cm}

The kernel $\ker (L(t)-\lambda)$, as in (0.10), contains two
distinguished (wave) functions, whose asymptotics is given later in
(2.5),
\be
\Phi^{(1)}(t,\lambda):=\frac{\BX_1(t,\lb)\tau(t)}{\tau(t)}\mbox{\,\,and\,\,}
\Phi^{(2)}(t,\lambda):=\frac{e^{\sum
t_i\lb^i}\BX_2(t,\lb)\tau(t)}{\tau(t)}
\ee
which Darboux transform, as follows:

\proclaim Theorem 0.2. The wave functions $\Phi^{(1)}$ and
$\Phi^{(2)}$ for the $L$-operator are Darboux transformed into wave
functions
$\tilde\Phi^{(1)}$ and $\tilde\Phi^{(2)}$ for the Darboux transformed
operator $\tilde L$; they are given by Wronskian
formulas, also expressible in terms of the new
$\tau$-function $\tilde\tau$, to wit
\begin{eqnarray*}
\tilde\Phi^{(1)}(t,z)&=&\frac{\BX_1(t,z)\tilde\tau}{\tilde\tau}=
\frac{1}{z}\frac{\{\Phi^{(1)}
(t,z),\Phi(t,\lb)\}}{\Phi(t,\lb)}=\frac{1}{z}L_+\Phi^{(1)}(t,z)\\
\tilde\Phi^{(2)}(t,z)&=&\frac{e^{\xi(t,z)}
\BX_2(t,z)\tilde\tau}{\tilde\tau}=
\frac{z}{\lb- z}\frac{\{\Phi^{(2)}
(t,z),\Phi(t,\lb)\}}{\Phi(t,\lb)}=\frac{z}{\lb-z}L_+\Phi^{(2)}(t,z),
\end{eqnarray*}
thus satisfying
$$
\tilde L\tilde\Phi^{(i)}=z\tilde\Phi^{(i)}\mbox{\,\,and\,\,}
\frac{\pl\tilde\Phi^{(i)}}{\pl t_n}=(\tilde
L^n)_+\tilde\Phi^{(i)},\quad i=1,2.
$$

\proclaim Theorem 0.3. The following holds:
$$
\mbox{Toda for $L$ and $\tilde
L\DF$ KM-lattice for $L_-$ and $L_+$,} .
$$
In coordinates, the latter takes theform
\be
\dot\al_n=(\beta_{n+1}-\beta_n)\al_n,\quad\dot\beta_n=(\al_n-\al_{n-1})
\beta_n;
\ee
moreover $\al_n$ and $\beta_n$ satisfy the Ricatti
equations, with coefficients given by the entries of $L$,
\begin{eqnarray*}
\dot\al_n&=&-\al^2_n+(b_{n+1}-\lb)\al_n-a_n\\
\dot\beta_n&=&\beta^2_n-(b_n-\lb)\beta_n+a_n.
\end{eqnarray*}
in well known analogy with the Sturm-Liouville situation.

\vspace{1cm}

Consider {\em band matrices} of the form
\be
L=\sum_{-p\leq i\leq p}a_i\Lb^i,\quad a_p=I,
\ee
with $a_i$ being diagonal matrices; define Toda lattice vector
fields, as follows:
$$
\frac{\pl L}{\pl x_i}=[(\overline{L^{i/p}})_+,L],\quad
\frac{\pl L}{\pl
y_i}=[(\underline{L^{i/p}})_-,L],\mbox{\,\,for\,\,}i=1,2,...,p\not| i
$$
\be
\frac{\pl L}{\pl t_{ip}}=[(L^i)_+,L],\quad i=1,2,...\,.
\ee
Note $\overline{L^{i/p}}$ and $\underline{L^{i/p}}$ involve {\em right}
$p^{{\rm th}}$ roots and {\em left} $p^{{\rm th}}$ roots:
\bea
\overline{L^{i/p}}&=&(\overline{L^{1/p}})^i=\left(\Lb+\sum_{k\leq
0}b_k\Lb^k\right)^i\nonumber\\
\underline{L^{i/p}}&=&(\underline{L^{1/p}})^i=\left(c_{-1}\Lb^{-1}+
\sum_{k\geq 0}c_k\Lb^k\right)^i;
\eea
with this notation, the vector fields (0.16) preserve the
band structure of $L$.

Then $L$ can be expressed in terms of a string of $\tau$-functions,
\be
\tau_n:=\tau_n(\hat x,\hat y,\hat t),
\ee
with $\hat x, ~\hat y,$ and $\hat t$ having certain components omitted:
\be
\hat x=(x_1,...,\hat x_p,...,\hat x_{2p},...),~~\hat y=(y_1,...,\hat
y_p,...,\hat y_{2p},...),~~\hat t=(t_p,t_{2p},t_{3p},...).
\ee
Define the following vertex operators:
\begin{eqnarray*}
\BX_1(\lb)&=&\chi(\lb)e^{\sum_1^{\iy}t_{pi}\lb^{pi}}e^{-\sum_1^{\iy}
\frac{\lb^{-pi}}{pi}\frac{\pl}{\pl t_{pi}}}e^{\sum_{p\not|
i}x_i\lb^i}e^{-\sum_{p\not| i}\frac{\lb^{-i}}{i}\frac{\pl}{\pl x_i}}\\
\BX_2(\lb)&=&\chi(\lb^{-1})e^{-\sum_1^{\iy}t_{pi}\lb^{pi}}
e^{\sum_1^{\iy}\frac{\lb^{-pi}}{pi}\frac{\pl}{\pl t_{pi}}}
e^{\sum_{p\not| i}y_i\lb^i}e^{-\sum_{p\not|
i}\frac{\lb^{-i}}{i}\frac{\pl}{\pl y_i}}.
\end{eqnarray*}

\proclaim Theorem 0.4. Each element of the 2$p$-dimensional
null-space\footnote{$\om$ is a primitive $p$th root of unity.}
$$
\ker(L-\lb^p)=\left\{\Phi(\lb)=\frac{\tilde\tau}{\tau}=
\frac{\displaystyle{\sum_{k=0}^{p-1}}\left(a_k\BX_1(\om^k\lb)+b_k
e^{\sum_{1}^{\iy}t_{ip}\lb^{ip}}\BX_2(\om^k\lb)\right)\tau}
{\tau}\right\},
$$
where $\Phi(\lb)$ satisfies
$$
L\Phi=\lb^p\Phi
$$
$$\frac{\pl\Phi}{\pl x_i}=(L^{i/p})_+\Phi,\quad
\frac{\pl\Phi}{\pl
y_i}=(L^{i/p})_-\Phi,\quad\mbox{for\,\,}i=1,2,...\mbox{\,\,with\,\,}p\not|
i
$$
\be
\frac{\pl\Phi}{\pl t_{ip}}=(L^i)_+\Phi,\quad\mbox{for\,\,}i=1,2,...,
\ee
determines a Toda-Darboux transform
$$
L-\lb^p\longmapsto (\beta\Lb^0+\Lb)(L-\lb^p)(\beta\Lb^0+\Lb)^{-1}
$$
with
$$
\beta=-\frac{\Lb\Phi(\lb)}{\Phi(\lb)};
$$
it acts on $\tau$ as
$$
\tau\longmapsto\tilde\tau=\tau\Phi=\displaystyle{\sum_{k=0}^{p-1}}\left(a_k\
BX_1(\om^k\lb)+b_k
e^{\sum_{1}^{\iy}t_{ip}\lb^{ip}}\BX_2(\om^k\lb)\right)\tau.
$$

For a broad account of Darboux transforms in a variety of
situations, see the book of Matveev and Salle \cite{MS}, which also
contains a very extensive bibliography. Darboux transforms for
differential operators, and their connections with
$\tau$-functions have been investigated in \cite{F,AvM1,vM} for 2nd
order differential operators and by Bakalov, Horozov and Yakimov
\cite{BHY1} for general differential operators. In fact, using our
vertex operator methods (see \cite{AvM1}), the latter results can be
made quite a bit more precise. The connection with the KM-lattice was
first made in
\cite{A}.

\vspace{1cm}

\noindent Table of contents:
\medbreak

\noindent 1. The 2-Toda lattice
\newline \noindent 2. Reduction from 2-Toda to tridiagonal 1-Toda, and
vertex operators
\newline \noindent 3. Toda-Darboux transformations
\newline \noindent 4. Wronskians and Vertex operators for 1-Toda
\newline \noindent 5. Borel factorization, KM-lattice and Ricatti
equations
\newline \noindent 6. Toda flows and Darboux transforms for band
matrices

\medbreak

\section{The 2-Toda lattice}

\setcounter{equation}{0}

Consider the splitting
ofthe algebra ${\cal D}$ of pairs $(P_1,P_2)$ of
infinite ($\BZ \times \BZ$) matrices such that $(P_1)_{ij}=0$ for
$j-i\gg0$ and $(P_2)_{ij}=0$ for $i-j\gg0$, used in \cite{ASV1};
to wit:
\begin{eqnarray*}
\DR&=&\DR_u+\DR_{\ell},\\
\DR_u&=&\bigl\{(P,P)\bigm|P_{ij}=0\hbox{ if }|i-j|\gg0\bigr\}
=\bigl\{(P_1,P_2)\in\DR\bigm| P_1=P_2 \bigr\},\\
\DR_{\ell}&=&\bigl\{(P_1,P_2)\bigm|(P_1)_{ij}=0\hbox{ if }j\ge i,\
(P_2)_{ij}=0\hbox{ if }i>j\bigr\},
\end{eqnarray*}
with $(P_1,P_2)=(P_1,P_2)_u+(P_1,P_2)_{\ell}$ given by
\begin{equation}
\setcounter{equation}{1}
\begin{array}{c}
(P_1,P_2)_u=(P_{1+}+P_{2-},P_{1+}+P_{2-}),\\
[3mm]\nonumber\\
(P_1,P_2)_{\ell}=(P_{1-}-P_{2-},P_{2+}-P_{1+});\nonumber
\end{array}
\end{equation}
$P_+$ and $P_-$ denote the upper (including diagonal)
and strictly lower triangular parts of the matrix $P$, respectively.


The two-dimensional Toda lattice equations
\begin{equation}
{\pl L\over\pl x_n}=\bigl[\bigl(L_1^n,0\bigr)_u,L\bigr]
\quad\hbox{and}\quad
{\pl L\over\pl y_n}=\bigl[\bigl(0,L_2^n\bigr)_u,L\bigr]
\quad n=1,2,\dots
\end{equation}
are deformations of a pair of infinite matrices
\begin{equation}
L=(L_1,L_2)
=\left(\sum_{-\iy<i\le1}a^{(1)}_i\Lambda^i,\sum_{-1\le
i<\iy}a^{(2)}_i\Lambda^i\right)
\in\DR,
\end{equation}
where $a_i^{(1)}$ and $a_i^{(2)}$ are diagonal matrices depending on
$x=(x_1,x_2,\dots)$ and $y=(y_1,y_2,\dots)$, such that
$$
a_1^{(1)}=I\quad\hbox{and}\quad \bigl(a_{-1}^{(2)}\bigr)_{nn}\ne
0\quad\mbox{for all}\,\, n.
$$

In their 2-Toda theory, Ueno-Takasaki \cite{UT} also introduce a
pair of wave vectors
$\Psi=(\Psi_1,\Psi_2)$, satisfying $(L_1,L_2)\Psi=(z,z^{-1})\Psi$
and\footnote{Here the action
is viewed componentwise, e.g., $(A,B)\Psi=(A\Psi_1,B\Psi_2)$ or
$(z,z^{-1})\Psi=(z\Psi_1,z^{-1}\Psi_2)$.}
\begin{equation}
\left\{
\begin{array}{l}
\frac{\pl }{\pl x_n}\Psi=(L_1^n,0)_u \Psi=((L_1^n)_+,(L_1^n)_+)\Psi\\
\frac{\pl }{\pl y_n}\Psi=(0,L_2^n)_u
\Psi=((L_2^n)_-,(L_2^n)_-)\Psi
 \end{array}
\right.\label{DE7}
\end{equation}
In \cite{UT}, it is shown that the wave vectors $\Psi$ can be
expressed in terms of one sequence of
$\tau$-functions $\tau(n,t,s)=
\tau_n(t_1,t_2,\dots;s_1,s_2,\dots),\quad n\in\BZ$,
to wit:
\be
\Psi_1(x,y;z)=\biggl(
        \frac{\tau_n(x-[z^{-1}],y)}{\tau_n(x,y)}
e^{\sum^{\iy}_1 x_iz^i}z^n
\biggr)_{n\in\BZ},
\ee
\be
\Psi_2(x,y;z)=\biggl(
        \frac{\tau_{n+1}(x,y-[z])}{\tau_n(x,y)}
e^{\sum^{\iy}_1 y_iz^{-i}}z^n
\biggr)_{n\in\BZ},
\ee
with $\tau$ satisfying the following bilinear identities:
$$
\oint_{z=\iy}\tau_n(x-[z^{-1}],y)\tau_{m+1}(x'+[z^{-1}],y')
e^{\sum_1^{\iy}(x_i-x'_i)z^i}
z^{n-m-1}dz
$$
\be
=\oint_{z=0}\tau_{n+1}(x,y-[z])\tau_m(x',y'+[z])
e^{\sum_1^{\iy}(y_i-y'_i)z^{-i}}z^{n-m-1}dz.
\ee
for all $m,n \in \BZ$. Conversely, any $\tau$-vector satisfying the
bilinear identity (1.7) leads to a solution of the 2-Toda lattice; see
\cite{TT}.

Upon introducing appropriate shifts of $x$ and $y$, and
evaluating the contour integration over a contour
about $\iy$ and the singularities, created by the shifts,
one finds the following Fay identities, due to \cite{AvM3}; they
will be useful later:
$$
\tau_n(x-[z^{-1}],y+[v]-[u])\tau_n(x,y)-\tau_n(x,y+[v]-[u])\tau_n(x-
[z^{-1}],y)
$$
\be
=\frac{v-u}{z}\tau_{n+1}(x,y-[u])\tau_{n-1}(x-[z^{-1}],y+[v])
\ee
and
$$
\tau_n(x,y+[v_1])\tau_{n+1}(x+[z_1^{-1}]-[z_2^{-1}],y-[v_2])
\frac{z_1^{-1}}{z_1^{-1}-z_2^{-1}}\hspace{3cm}
$$
$$\hspace{3cm}+\,\tau_n(x+[z_1^{-1}]-[z_2^{-1}],y+[v_1])
\tau_{n+1}(x,y-[v_2])\frac{z_2^{-1}}{z_2^{-1}-z_1^{-1}}
$$

$$=\tau_{n+1}(x+[z_1^{-1}],y)\tau_n(x-[z_2^{-1}],y+[v_1]-[v_2])
\frac{v_1}{v_1-v_2}\hspace{3cm}
$$
\be
\hspace{3cm}+\,\tau_{n+1}(x+[z_1^{-1}],y+[v_1]-[v_2])\tau_n
(x-[z_2^{-1}],y)\frac{v_2}{v_2-v_1}.
\ee

Consider the following 2-Toda vertex
operators\footnote{$\chi(\lb)=(\lb^n)_{n\in\BZ}$}
$$
\BY_1(\lb)=\chi(\lb)X(x;\lb)\quad\mbox{and}\quad\BY_2(\mu)=\chi(\mu)
X(y,\mu^{-1})\Lambda,
$$
where
$
X(x,\lb)$ is
given by
(0.8).
In \cite{AvM3}, we have shown that for fixed $\lb,\mu\in\BC$, the
vertex operator
$$
a\BY_1(\lb)+b\BY_2(\mu)
$$
maps 2-Toda $\tau$-vectors into themselves. Spelled out,
\medbreak

\noindent $\Biggl(\Bigl(a\BY_1(\lb)+b\BY_2(\mu)\Bigr)\tau\Biggr)_n$
\be
=a e^{\sum_1^{\iy}x_i\lb^i}\lb^n\tau_n(x-[\lb^{-1}],y)
+b e^{\sum_1^{\iy}y_i\mu^{-i}}\mu^n\tau_{n+1}(x,y-[\mu])
\ee
is a new $\tau$-vector for the 2-Toda lattice.

\section{Reduction from 2- to tridiagonal 1-Toda, and vertex operators}
\setcounter{equation}{0}

In the notation of section 1, consider the locus of $(L_1,L_2)$'s in
$\DR_u$, i.e., such that $L_1=L_2$. Since the equations (1.2) imply
\be
{\pl L_i\over\pl x_n}=\bigl[\bigl(L_1^n\bigr)_+,L_i\bigr]
\quad\hbox{and}\quad
{\pl L_i\over\pl y_n}=\bigl[\bigl(L_2^n\bigr)_-,L_i\bigr]
\quad n=1,2,\dots
\ee
we have that along $\DR_u$
\be
\frac{\pl(L_1-L_2)}{\pl x_n}=0\quad\quad\quad\frac{\pl(L_1-L_2)}{\pl
y_n}=0,
\ee
implying that the vector fields $\displaystyle{\frac{\pl}{\pl t_n}}$
and
$\displaystyle{\frac{\pl}{\pl s_n}}$ are tangent to
$\DR_u$. Also when $L_1=L_2$, the matrix $L:=L_1=L_2$ is tridiagonal.
Moreover
\begin{equation}
\left(\frac{\pl}{\pl
x_n}+\frac{\pl}{\pl
y_n}\right)L_1=[(L_1^n)_++(L_2^n)_{-},L_1]=
[(L_1^n)_++(L_1^n)_{-},L_1]=0;
\ee
setting
$$
x_n=\frac{t_n+s_n}{2},\quad\quad y_n=\frac{-t_n+s_n}{2}
\quad\mbox{ and }\quad t_n=x_n-y_n,\quad\quad s_n=x_n+y_n,$$
with
\be
\frac{\pl}{\pl t_n}=\frac{1}{2}\left(\frac{\pl}{\pl
x_n}-\frac{\pl}{\pl y_n}\right),\quad\quad
\frac{\pl}{\pl s_n}=\frac{1}{2}\left(\frac{\pl}{\pl
x_n}+\frac{\pl}{\pl y_n}\right),
\ee
equation (2.1) implies that $L=L_1=L_2$ is independent
of
$s$.

The 2-Toda lattice wave functions $\Psi_1$ and $\Psi_2$, properly
reduced, yield two distinguished eigenfunctions for the 1-Toda
lattice; they have the following expressions in terms of the 1-Toda
$\tau$-functions:
\begin{eqnarray}
\Phi^{(1)}(t,z)&:=&\Psi_1(x,y;z)e^{-\sum_1^{\iy}y_iz^i}
\nonumber\\
 & & \nonumber\\
&=&e^{\sum_1^{\iy}t_iz^i}\left(z^n\frac{\tau_n(t-[z^{-1}])}{\tau_n(t)}
\right)_{n\in\BZ} \nonumber\\
&=& \frac{\BX_1(z)\tau}{\tau} \nonumber\\
& & \nonumber\\
&=& e^{\sum_1^{\iy}t_iz^i}z^n
\left(1+O(z^{-1})\right)\\
& &\nonumber\\
\Phi^{(2)}(t,z)&:=&\Psi_2(x,y;z^{-1})e^{-\sum_1^{\iy}y_i
z^i}\nonumber\\ & &
\nonumber\\ &=&
\left(z^{-n}\frac{\tau_{n+1}(t+[z^{-1}])}
{\tau_n(t)}\right)_{n\in\BZ}\nonumber\\
&=& \frac{e^{\sum_1^{\iy}t_iz^i}
\BX_2(z)\tau}{\tau}\nonumber\\
 \nonumber\\
&=&
z^{-n}\left(\frac{\tau_{n+1}(t)}{\tau_n(t)}+O(z^{-1})\right)
,
\end{eqnarray}
in terms of the vertex operators, already defined in (0.7), namely
\be
\BX_1(z)=\chi(z)X(t,z)\quad\mbox{and}\quad
\BX_2(z)=\chi(z^{-1})
X(-t,z)\Lambda;
\ee
also, for later use, recall from (0.4),
\be
\BX(t,z):=\Lambda^{-1}\chi(z^2)e^{\sum t_i z^i}e^{-2\sum
\frac{z^{-i}}{i}
\frac{\pl}{\pl t_i}};
\ee
Using (1.4), (2.2) and the fact that $L_1=L_2$, one checks
$$
L\Phi^{(1)}=L_1\Phi^{(1)}=z\Phi^{(1)}\quad\quad
L\Phi^{(2)}=L_2\Phi^{(2)}=z\Phi^{(2)}
$$
and
\begin{eqnarray}
\frac{\pl\Phi^{(1)}}{\pl t_n}&=&\frac{1}{2}\left(\frac{\pl}{\pl x_n}
-\frac{\pl}{\pl y_n}
\right)\Psi_1(x,y;z)
e^{-\sum_1^{\iy}y_iz^i}=(L^n)_+\Phi^{(1)}\\
\frac{\pl\Phi^{(2)}}{\pl t_n}&=&\frac{1}{2}\left(\frac{\pl}{\pl x_n}
-\frac{\pl}{\pl y_n}
\right) \Psi_2(x,y;z^{-1})
e^{-\sum y_iz^i}=(L^n)_+\Phi^{(2)}.
\end{eqnarray}
Therefore
\be
\Phi(t,z)=a\Phi^{(1)} + b \Phi^{(2)}
\ee
is the most general solution to the following problem
\be
L\Phi=z\Phi\quad\mbox{and}\quad\frac{\pl\Phi}{\pl t_n}=(L^n)_+\Phi,
\ee
with
\be
\frac{\pl L}{\pl t_n}=[(L^n)_+,L].
\ee

\proclaim Lemma 2.1. If $(\tau_n)_{n\in\BZ}$ is a $\tau$-vector for
1-Toda, then for arbitrary $z\in\BC^*$, $a,b\in\BC$, the
vectors\footnote{$
\xi(t,z):=\sum^{\iy}_1 t_i z^i.$}
\be
\left((a\BX_1(t,z)+b e^{\xi(z)}\BX_2(t,z))\tau\right)_{n\in\BZ}
\ee
and
\be
\left((1+c\BX(t,z))\tau\right)_{n \in \BZ}
\ee
are 1-Toda vectors. The vertex operators $\BX_1(y)$ and
$e^{\xi(t,z)}\BX_2(z)$ satisfy the commutation relation
\be
\frac{y}{1-y/z}\BX_1(y)e^{\xi(z)}\BX_2(z)=e^{\xi(z)}\BX_2(z)\BX_1(y).
\ee

\proof According to the reduction above from 2- to 1-Toda, the
tau-functions $\tau_n$ are independent of the sum of the arguments.
Thus, we may write $\tau_n(x-y)=\tau_n(t)$ for $\tau_n(x,y)$. Set
$\lb=\mu^{-1}=z\in\BC^*$ in
(1.10); moreover, it is legitimate to multiply the
$\tau$-vector with an exponential in $y_i$ with constant coefficients.
Therefore the following expression is a 1-Toda $\tau$-vector:
\medbreak
\noindent
$\displaystyle{e^{-\sum_1^{\iy} y_iz^i}\left(\left(a\BY_1(z)
+b\BY_2(z^{-1})\right)\tau
\right)_n}$
\begin{eqnarray*}
&=&e^{-\sum y_iz^i}\left(ae^{\sum_1^{\iy}x_iz^i}z^n\tau_n(x-y-
[z^{-1}])+be^{\sum y_iz^i}z^{-n}\tau_{n+1}
(x-y+[z^{-1}])\right)\\
&=&ae^{\sum
t_iz^i}z^n\tau_n(t-[z^{-1}])+bz^{-n}\tau_{n+1}(t+[z^{-1}])\\
&=&\left((a\BX_1(z)+be^{\xi(z)}\BX_2(z))\tau\right)_n.
\end{eqnarray*}
Finally, applying the inverse of $e^{\xi(y)}\BX_2(y)$ to the above
equation and taking a limit, when $y\rightarrow z$, leads to the result
(2.15), after noting that
$$
\lim_{y\rightarrow z}\left(e^{\xi(y)}\BX_2(y)\right)^{-1}
\left(e^{\xi(z)}\BX_2(z)\right)=\lim_{y\rightarrow
z}\Lambda^{-1}\chi(yz^{-1})e^{\sum \frac{z^{-i}-y^{-i}}{i}
\frac{\pl}{\pl t_i}}\Lambda=I
$$

$$\lim_{y\rightarrow z}
\frac{1}{1-z/y}\left(e^{\xi(y)}\BX_2(y)\right)^{-1}\BX_1(z)
\hspace{6cm}
$$
\begin{eqnarray*}
\hspace{3cm} &=&\lim_{y\rightarrow z}
\frac{1}{1-z/y}\Lambda^{-1}\chi(y)\chi(z)e^{-\sum
\frac{y^{-i}}{i}
\frac{\pl}{\pl t_i}}  e^{\sum t_i z^i} e^{-\sum \frac{z^{-i}}{i}
\frac{\pl}{\pl t_i}}\\
\hspace{3cm} &=&\Lambda^{-1}\chi(z^2)e^{\sum t_i z^i}e^{-2\sum
\frac{z^{-i}}{i}
\frac{\pl}{\pl t_i}}\\
&=&\BX(t,z),
\end{eqnarray*}
using the commutation relation
$$
[e^{-\sum \frac{y^{-i}}{i}
\frac{\pl}{\pl t_i}},e^{\sum t_i z^i}]=-\frac{z}{y}
e^{\sum t_i z^i}e^{-\sum \frac{y^{-i}}{i} \frac{\pl}{\pl t_i}}.
$$
\hfill \qed

\section{Toda-Darboux transformations}
\setcounter{equation}{0}

The purpose of this section is to establish theorem 0.1, which we
restate in an explicit form.

\proclaim Proposition 3.1. For $L$ evolving according to the Toda
lattice equations (2.9), the Borel decomposition
$L(t)-\lb=L_-(t)L_+(t)$ is given for all $t\in \BC^{\iy}$, by
\be
\al_n=\frac{\pl}{\pl
t_1}\log\left(e^{-\sum t_i \lb^i}\Phi_{n+1}(t,\lb)\right)
=\frac{\pl}{\pl t_1}\log\Phi_{n+1}(t,\lb)-\lb
\ee
\be
\beta_n=-\frac{\pl}{\pl
t_1}\log\left( \frac{\tau_n}{\tau_{n+1}}\Phi_n(t,\lb)
\right)=-\frac{\Phi_{n+1}(t,\lb)}{\Phi_n(t,\lb)}
\ee
in terms of the vector $\Phi$, defined in (2.11), i.e.,
$$
L-\lambda=L_-L_+=\left(\Lambda^{-1}(\Lambda\frac{\pl}{\pl t_1}\log
\Phi(t,\lambda )-\lambda)+\Lambda^0 \right)
\left( -\left(
\frac{\Lambda\Phi(t,\lambda}{\Phi(t,\lambda)}\right)\Lambda^0+\Lambda
\right);
$$
so, $\Phi$ belongs to the kernel of $L_+$, i.e.,
$$
L_+\Phi=0.
$$
Finally the Darboux map
$$
L(t)-\lb=L_-(t)L_+(t)\mapsto\tilde L(t)-\lb=L_+(t)L_-(t)
$$
is Toda-Darboux, i.e., $\tilde L(t)$ satisfies the Toda
lattice as well.

\proof Let us begin by showing the second identity in (3.2), using
(0.3) and (2.12):
$$
\Phi_n \frac{\pl}{\pl
t_1}\log\left( \frac{\tau_n}{\tau_{n+1}}\Phi_n(t,\lb)
\right)=-b_n\Phi_n+\frac{\pl \Phi_n}{\pl
t_1}=-b_n\Phi_n+(L_+\Phi)_n=\Phi_{n+1}.
$$
To establish the theorem, we must check, at first, that $L-\lb=L_-L_+$,
with the entries $\al_n$ and $\beta_n$, given by (3.1) and (3.2);
secondly, we must check that $\tilde L-\lb=L_+L_-$ evolves according to
the Toda lattice.

To begin with, we verify the transformation (0.6):
$b_n-\lb=\al_{n-1} + \beta_n$ and $a_{n-1}=\al_{n-1} \beta_{n-1}$,
with $\al_n,\beta_n,$ and $b_n$ equal to (3.1),(3.2) and (0.3)
respectively:
$$\al_{n-1}+\lb +\beta_n-b_n=\frac{\pl}{\pl t_1}
\left( \log \Phi_n(t,\lb) - \log \frac{\tau_n}{\tau_{n+1}}\Phi_n(t,\lb)
-\log \frac{\tau_{n+1}}{\tau_n}\right)=0
$$
\begin{eqnarray*}
\Phi_{n-1}(a_{n-1}-\al_{n-1} \beta_{n-1})&=&
     \Phi_{n-1}\left(a_{n-1}+\left(\frac{\pl}{\pl
t_1}\log\Phi_n-\lb\right)\frac{\Phi_n}{\Phi_{n-1}}\right)\\
&=& a_{n-1}\Phi_{n-1}+\frac{\pl \Phi_n}{\pl t_1}-\lb\Phi_n\\
&=&\left(\left(\frac{\pl}{\pl t_1}-L+L_-\right)\Phi \right)_n=0.
\end{eqnarray*}
Remember from (0.6), that the Darboux map
\be
L-\lb=L_-L_+\mapsto\tilde L-\lb=L_+L_-
\ee
reads componentwise:
$$
b_n-\lb=\al_{n-1}+\beta_n\longmapsto\tilde
b_n-\lb=\al_n+\beta_n=(b_n-\lb)+ (\al_n-\al_{n-1})
$$
$$
a_{n-1}=\al_{n-1}\beta_{n-1}\longmapsto\tilde
a_{n-1}=\al_{n-1}\beta_n=a_{n-1}\frac{\beta_n}{\beta_{n-1}}
$$
with $\al_n$ and $\beta_n$ given by (3.2). So, in terms of $\tau_n$ and
$\Phi_n$, we have
\begin{eqnarray*}
\tilde b_n-\lb&=&(b_n-\lb)+(\al_n-\al_{n-1})\\
&=&\frac{\pl}{\pl t_1}\log\frac{\tau_{n+1}}{\tau_n}-\lb+
\frac{\pl}{\pl t_1}\log\frac{\Phi_{n+1}(t,\lb)}{\Phi_n(t,\lb)}\\
&=&\frac{\pl}{\pl
t_1}\log\frac{\tau_{n+1}\Phi_{n+1}(t,\lb)}{\tau_n\Phi_n(t,\lb)}-\lb\\
&=&\frac{\pl}{\pl t_1}\log\frac{\tilde\tau_{n+1}}{\tilde\tau_n}-\lb\\
\tilde a_{n-1}&=&a_{n-1}\frac{\beta_n}{\beta_{n-1}}\\
&=&\frac{\tau_{n-1}\tau_{n+1}}{\tau^2_n}\cdot
\frac{\Phi_{n-1}(t,\lb)\Phi_{n+1}(t,\lb)}{\Phi_n(t,\lb)^2}\\
&=&\frac{\tilde\tau_{n-1}\tilde\tau_{n+1}}{\tilde\tau_n^2},
\end{eqnarray*}
where
$$
\tilde\tau=\tau(t)\Phi(t,\lb)=\left(a\BX_1(\lb)+be^{\xi(\lb)}\BX_2(\lb)
\right)\tau
$$
is a new $\tau$-vector, by virtue of Lemma 2.1. Hence, the matrix
$\tilde L$, parametrized by this new $\tau$-vector $\tilde \tau$,
satisfies the symmetric $1$-Toda equations. \qed

\section{Wronskians and Vertex operators for 1-Toda}

The aim of this section is to prove Theorem 0.2. The Wronskian
on vectors
$f=(f_n)_{n\in\BZ}$ was defined in (0.9), the vertex operators
$\BX_1(t,y)$ and $\BX_2(t,z)$ in (0.7), and
$\xi(t,z)=\xi(z)=\sum^{\iy}_1 t_i z^i$.

\setcounter{equation}{0}

\proclaim Proposition 4.1. The following identities hold:
\begin{eqnarray*}
\{\Phi^{(1)}(t,y),\Phi^{(1)}(t,z)\}&=&y\frac{\BX_1(y)\BX_1(z)\tau}{\tau}\\
\{\Phi^{(1)}(t,y),\Phi^{(2)}(t,z)\}&=&y\frac{\BX_1(y)
e^{\xi(z)}\BX_2(z)\tau}{\tau}=\left(1-\frac{y}{z}\right)
\frac{e^{\xi(z)}\BX_2(z)\BX_1(y)\tau}{\tau}\\
\{\Phi^{(2)}(t,y),\Phi^{(2)}(t,z)\}&=&-\left(1-\frac{z}{y}
\right)\frac{e^{\xi(y)} \BX_2(y)e^{\xi(z)}\BX_2(z)\tau}{\tau}.
\end{eqnarray*}

\proof Using, in the third equality, Fay identity (1.9) with
the shift $t\mapsto t-[z_1^{-1}]$, and the limits $z_1\rightarrow
z$,
$z_2\rightarrow y$ and $v_1=v_2\rg 0$,
\medbreak
\noindent$\displaystyle{\{\Phi^{(1)}_n(t,y),\Phi_n^{(1)}(t,z)\}}$
\begin{eqnarray*}
&=&\frac{e^{\xi(y)+\xi(z)}}{\tau_n\tau
_{n+1}}(yz)^n\\
& &\hspace{1cm}
\left(y\tau_{n+1}(t-[y^{-1}])\tau_n(t-[z^{-1}])-z\tau_n(t-
[y^{-1}])\tau_{n+1}(t-[z^{-1}])\right)\\
&=&\frac{e^{\xi(y)+\xi(z)}}{\tau_n\tau_{n+1}}(yz)^n(y-z)\tau_{n+1}\tau_n(
t-[y^{-1}]-[z^{-1}])\\
&=&e^{\xi(y)+\xi(z)}(y-z)(yz)^n\frac{\tau_n(t-[y^{-1}]
-[z^{-1}])}{\tau_n(t)}\\
&=&y\left(\frac{\BX_1(y)\BX_1(z)\tau}{\tau}\right)_n.
\end{eqnarray*}
\medbreak

Using in the third equality Fay identity (1.8) with $z\rg y$,
$u\rg z^{-1}$, $v\rg 0$ and $n\mapsto n+1$:
\medbreak
\noindent$\displaystyle{\{\Phi^{(1)}_n(t,y),\Phi_n^{(2)}(t,z)\}}$
\begin{eqnarray*}
&=&\frac{e^{\xi(y)}}{\tau_n\tau_{n+1}}
\left(\frac{y}{z}\right)^n\\
& & \hspace {1cm}\left(y\tau_{n+1}(t-[y^{-1}])\tau_{n+1}(
t+[z^{-1}])-z^{-1}\tau_n(t-[y^{-1}])\tau_{n+2}(t+[z^{-1}])\right)\\
&=&\frac{e^{\xi(y)}}{\tau_n\tau_{n+1}}\left(\frac{y}{z}\right)^n
y\tau_{n+1}(t-[y^{-1}]+[z^{-1}])\tau_{n+1}\\
&=&e^{\xi(y)}y\left(\frac{y}{z}\right)^n
\frac{\tau_{n+1}(t-[y^{-1}]+[z^{-1}])}{\tau_n}\\
&=&y\left(\frac{\BX_1(y)
e^{\xi(z)}\BX_2(z)\tau}{\tau}\right)_n=\left(1-\frac{y}{z}\right)\left(
\frac{e^{\xi(z)}\BX_2(z)\BX_1(y)\tau}{\tau}\right)_n,
\end{eqnarray*}
using the commutation relation (2.16), in the last equality.

\medbreak
Using in the third equality the same
identity as for the first wronskian, but with $t\mapsto
t+[y^{-1}]+[z^{-1}]$ and
$n\mapsto n+1$:
\medbreak
\noindent$\displaystyle{\{\Phi^{(2)}_n(t,y),\Phi_n^{(2)}(t,z)\}}$
\begin{eqnarray*}
&=&\frac{1}{\tau_n\tau_{n+1}}
(yz)^{-n-1}\\
& & \hspace{1cm}\left(z\tau_{n+2}(t+[y^{-1}])\tau_{n+1}(
t+[z^{-1}])-y\tau_{n+1}(t+[y^{-1}])\tau_{n+2}(t+[z^{-1}])\right)\\
&=&-\frac{1}{\tau_n\tau_{n+1}}(yz)^{-n-1}(y-z)\tau_{n+2}
(t+[y^{-1}]+[z^{-1}])\tau_{n+1}\\
&=&-(y-z)(yz)^{-n-1}\frac{\tau_{n+2}(t+[y^{-1}]+[z^{-1}])}{\tau_n}\\
&=&-\left(1-\frac{z}{y}\right)
\left(\frac{e^{\xi(y)}\BX_2(y)e^{\xi(z)}\BX_2(z)\tau}{\tau}\right)_n
\end{eqnarray*} \qed

\medbreak
\underline{\sl Proof of Theorem 0.2}: Proposition 4.1 implies the
following relations:
$$
\{\Phi_n^{(1)}(t,z),\Phi_n(t,\lb)\}=z\left(
\frac{\BX_1(z)(a\BX_1(\lb)+be^{\xi(\lb)}\BX_2(\lb))\tau}{\tau}
\right)_n = z \frac{X_1(z)\tilde \tau}{\tau},
$$
and
\begin{eqnarray*}
\{\Phi_n^{(2)}(t,z),\Phi_n(t,\lb)\}&=& -(1-\frac{\lb}{z}) \left(
\frac{e^{\xi(z)}\BX_2(z)(a\BX_1(\lb)+be^{\xi(\lb)}\BX_2(\lb))\tau}{\tau}
\right)_n \\
&=& \frac{\lambda-z}{z} \frac{e^{\xi(z)} X_2(z) \tilde
\tau}{\tau},
\end{eqnarray*}
from which the proof follows.

\hfill\qed

\section{Borel factorization, KM-lattice and Ricatti equations}

\setcounter{equation}{0}
This section concerns itself with the proof of Theorem 0.3. For a fixed
$\lb\in\BC^*$, consider the Darboux map
\be
L-\lambda=L_-L_+\longmapsto\tilde L-\lambda=L_+L_-,
\ee
which in coordinates reads
\begin{eqnarray}
b_n-\lambda &=&\al_{n-1}+\beta_n\longmapsto\tilde
b_n-\lambda =\al_n+\beta_n=(b_n-\lambda)+ (\al_n-\al_{n-1})\nonumber\\
a_{n-1}&=&\al_{n-1}\beta_{n-1}\longmapsto\tilde
a_{n-1}=\al_{n-1}\beta_n=a_{n-1}\frac{\beta_n}{\beta_{n-1}}.
\end{eqnarray}
Remember the first Toda flow $(=\pl/\pl t_1=\, ^{\prime})$
\be
 b_n'=a_n-a_{n-1},\quad a_n'=a_n(b_{n+1}-b_n)
\ee
and the Kac-Moerbeke lattice, refered to with ${}^.~$:
\be
\dot\al_n=(\beta_{n+1}-\beta_n)\al_n,\quad\dot\beta_n=(\al_n-
\al_{n-1})\beta_n.
\ee

\proclaim Proposition 5.1. Toda for $L$ and $\tilde L\DF {\rm KM}$ for
$L_-$ and $L_+$.

\proof Assuming the KM-vector field on $(\al,\beta)$, one computes
\be
(b_n-\lb)^{\cdot}=\dot\al_{n-1}+\dot\beta_n=\al_n\beta_n-\al_{n-1}
\beta_{n-1}=a_n-a_{n-1}=(b_n-\lb)'
\ee
\be
\dot a_n=\dot\al_n\beta_n+\al_n\dot\beta_n=\al_n\beta_n(\beta_{n+1}-
\beta_n+\al_n-\al_{n-1})=a_n(b_{n+1}-b_n)=a_n'
\ee
and
\be
(\tilde
b_n-\lb)^{\cdot}=\dot\al_n+\dot\beta_n=\al_n\beta_{n+1}-\al_{n-1}
\beta_n=\tilde a_n-\tilde a_{n-1}=(\tilde b_n-\lb)'
\ee
\be
\dot{\tilde
a}_n=\dot\al_n\beta_{n+1}+\al_n\dot\beta_{n+1}=\al_n\beta_{n+1}(\beta_{n+1}
-\beta_n+\al_{n+1}-\al_n)=\tilde a_n(\tilde b_{n+1}-\tilde b_n)=\tilde
a_n'.
\ee
Conversely, assuming Toda on $L$ and $\tilde L$ leads to the following
four equations on $\al$ and $\beta$; the first two are (5.5) and
(5.8), with a shift, the last two are (5.6) and (5.7):
$$
\left\{
\begin{array}{ll}
\al_n'+\beta_{n+1}'&=\al_{n+1}\beta_{n+1}-\al_n\beta_n\\
 & \\
\al_n'\beta_{n+1}+\al_n\beta_{n+1}'&=\al_n\beta_{n+1}(\beta_{n+1}-
\beta_n+\al_{n+1}-\al_n)
\end{array}
\right.
$$
and$$
\left\{
\begin{array}{ll}
\al_n'\beta_n+\al_n\beta_n' &=\al_n\beta_n(\beta_{n+1}-
\beta_n+\al_n-\al_{n-1})\\
 & \\
\al_n'+\beta_n'&=\al_n\beta_{n+1}-\al_n\beta_n.
\end{array}
\right.
$$
Solving the first system in $\al_n'$ and $\beta_{n+1}'$ or the
second in $\al_n'$ and $\beta_n'$ leads to the KM-lattice
equations (5.4); so, now we may replace the differentiation ${}^.$
by ${}'~$.\qed

\vspace{1cm}

The KM-lattice equations (3.7)
$$
\al'_n=(\beta_{n+1}-\beta_n)\al_n\quad\mbox{and}\quad\beta'_n
=(\al_n-\al_{n-1})\beta_n,
$$
upon using $\beta_n=\frac{a_n}{\al_n}$, $\beta_{n+1}=b_{n+1}-\al_n-\lb$
in the first equation and
$\al_n=\frac{a_n}{\beta_n}$ and $\al_{n-1}=b_n-\lb-\beta_n$ in the
second equation, yield
\begin{eqnarray}
\al'_n&=&-\al^2_n+(b_{n+1}-\lb)\al_n-a_n\nonumber\\
\beta'_n&=&\beta^2_n-(b_n-\lb)\beta_n+a_n\quad\quad\mbox{(Ricatti
equations);}
\end{eqnarray}
the transformations
\be
\al_n=\frac{\pl}{\pl t_1}\log\ga_n~~~\mbox{and}~~~
\beta_n=-\frac{\pl}{\pl t_1}\log\vr_n
\ee
yield the second order linear equations
\be\ga''_n-(b_{n+1}-\lambda)\ga'_n+a_n\ga_n=0\ee
\be\vr''_n+(b_n-\lambda)\vr'_n+a_n\vr_n=0.\ee

\proclaim Proposition 5.2. The expressions
\be
\al_n=\frac{\pl}{\pl t_1}\log\ga_n:=\frac{\pl}{\pl
t_1}\log\left(e^{-\sum t_i \lb^i}\Phi_{n+1}(t,\lb)\right)
\ee
\be
\beta_n=-\frac{\pl}{\pl t_1}\log\vr_n:=-\frac{\pl}{\pl
t_1}\log\left( \frac{\tau_n}{\tau_{n+1}}\Phi_n(t,\lb)
\right),
\ee
given
by the 2-dimensional family $\Phi(t,z)=a\Phi^{(1)}+b\Phi^{(2)}$, provide
the most general solution to the Ricatti equations (5.9).

\proof Since the parametrization of $\al_n$ and $\beta_n$ in
Proposition 3.1 provide a Toda-Darboux transformation, then, by
proposition 5.1, $\al_n$ and $\beta_n$ satisfy the KM-Lattice, and
hence provide a 2-dimensional solution to the Ricatti equations
(5.9).

An {\em alternate proof} not using $\tau$-function theory proceeds
as follows: The Ricatti equations (5.9) are
equivalent via the transformation (5.2) and the Toda vector fields
(5.3). We check, for instance, that
$\vr_n$ satisfies (5.12); at first, we compute:
\be\Phi'_n=(L_+\Phi)_n=b_n\Phi_n+\Phi_{n+1}=\lb\Phi_n-a_{n-1}\Phi_{n-1},
\mbox{\,\,using\,\,}L\Phi=\lb \Phi.\ee
Using the first expression for $\Phi'_n$ and $\Phi'_{n-1}$ and the
Toda lattice equation for $a'_{n-1}$, we then find
\begin{eqnarray}
\Phi''_n&=&\lb\Phi'_n-a'_{n-1}\Phi_{n-1}-a_{n-1}\Phi'_{n-1}\nonumber\\
&=&\lb
(b_n\Phi_n+\Phi_{n+1})-a_{n-1}(b_n-b_{n-1})\Phi_{n-1}-a_{n-1}(b_{n-1}\Phi_{n-1}
+\Phi_n)\nonumber\\
&=&(b_n+\lb)\Phi_{n+1}+(b^2_n-a_{n-1})\Phi_n,\mbox{\,\,using\,\,}\lb
\Phi_{n}=(L\Phi)_{n}.
\end{eqnarray}
Then, using
\be\frac{\tau_{n+1}}{\tau_n}\left(
\frac{\tau_{n}}{\tau_{n+1}}\right)^{\prime}=-b_n\quad\mbox{ and }\quad
\frac{\tau_{n+1}}{\tau_n}\left(
\frac{\tau_n}{\tau_{n+1}}\right)^{\prime\prime}=-b'_n+b^2_n=a_{n-1}-a_n
+b^2_n,
\ee
we find
\medbreak
\noindent
$\displaystyle{\frac{\tau_{n+1}}{\tau_n}(\vr''_n+(b_n-\lb)
\vr'_n+a_n\vr_n)}$
\begin{eqnarray*}
&=&\Phi''_n-2b_n\Phi'_n+(-b'_n+b^2_n)\Phi_n+(b_n-\lb)
(\Phi'_n-b_n\Phi_n)+a_n\Phi_n\\
&\stackrel{\ast}{=}&\Phi''_n-(b_n+\lb)\Phi_{n+1}+(-b^2_n+a_{n-1})\Phi_n\\
&=&0,\quad\mbox{by (5.18)}
\end{eqnarray*}
in $\stackrel{\ast}{=}$, we have used (5.3), (5.17) and $\lb
\Phi=L\Phi$.
\qed

\section{Toda flows and Toda-Darboux transforms for band matrices}

\setcounter{equation}{0}

In this section, we factorize band matrices of the form
$L=\sum_{-p\leq i\leq p}a_i\Lb^i$, and study their Toda-Darboux
transforms.

\proclaim Lemma 6.1. Consider the difference operator
\be
L=a_{-r}\Lb^{-r}+a_{-r+1}\Lb^{-r+1}+...+a_{n-r-1}\Lb^{n-r-1}+\Lb^{n-r}
\ee
with $n\geq 2$, $r\geq 0$, diagonal matrices $a_j$, with leading term
$a_{-r}(j)\neq 0$ for
$j$ sufficiently small. Then any choice of basis
$\Phi^{(1)},...\Phi^{(n)}\in\ker L$ leads to a factorization
of \footnote{Wronskian
$=W_k(\ell)=W[\Phi^{(1)},\Phi^{(2)},...,\Phi^{(k)}](\ell)=\det\left(
\Phi^{(j)}_{i+\ell-1}\right)_{1\leq i,j\leq k}$. Note $\Lambda^{-1}$
in the formula (6.2) refers to the expression in brackets only!}
$L$:
\bea
L&=&\left(I-\Lb^{-1}(\Lb^{-r+1}\beta_n)\right)
\left(I-\Lb^{-1}(\Lb^{-r+2}\beta_{n-1})\right)...
\left(I-\Lb^{-1}(\beta_{n-r+1})\right)\cdot\nonumber\\
& &\quad\cdot\,
\left(\Lb-\beta_{n-r}I\right)\left(\Lb-\beta_{n-r-1}I\right)...\left(\Lb-\beta_1
I
\right),
\eea
with
$$
\beta_k=\frac{\Lb\al_k}{\al_k},~~\al_k(\ell)=\frac{W_k
(\ell)}{W_{k-1}(\ell)},~~W_k(\ell)=W[\Phi^{(1)},
\Phi^{(2)},...,\Phi^{(k)}](\ell).
$$

\proof \underline{Step 1}: $r=0$

\noindent First we prove the statement for $r=0$ by induction on the
degree of
$L$. Define the operator $L_i$,
$$
L_i(f):=\frac{W[f,\Phi^{(1)},...,\Phi^{(i)}]}
{W[\Phi^{(1)},...,\Phi^{(i)}]}
$$
which is the unique operator of the form (6.1) with $r=0$, $n=i$ such
that $\ker L_i=\{\Phi^{(1)},...,\Phi^{(i)}\}$. But clearly
$$
L_{i+1}(f)=\frac{W[L_i(f),L_i(\Phi^{(i+1)})]}{
L_i(\Phi^{(i+1)})}=\left(\Lb-\frac{\Lb\al_i}{\al_i}I\right)L_i(f),
$$
with $\al_i=L_i(\Phi^{(i+1)})=W_{i+1}/W_i$. So, by induction $L_i(f)$
factors, thus leading to (6.2) for $r=0$.

\medbreak

\noindent\underline{Step 2}: $r\neq 0$

The case $r\neq 0$ is taken care of by multiplying (6.2) to the left
with $\Lb^r$:
$$
\Lb^rL=(\Lb-\beta_nI)...(\Lb-\beta_1I),
$$
on which you apply step 1.\qed

\vspace{0.5cm}

In the next Lemma, we use definition (0.19) for $\hat x,\hat y,\hat
t$; also define
$$
\hat x-[\lb^{-1}]:=\left(x_i-\frac{\lb^{-i}}{i}\right)_{i\neq p},\hat
t-[\lb^{-1}]=\left(t_{ip}-\frac{\lb^{-ip}}{ip}\right)_{i=1,2,...}.
$$

\proclaim Lemma 2. On the locus $L:=L_1^p=L_2^p$, the 2-Toda vector
fields (2.1) on $(L_1,L_2)$ take the form (0.16), with
$\tau$-functions of the form (0.18) and with the 2$p$-dimensional
null-space $\ker(L-\lb^p)$ given by
\be
\Phi(\hat x,\hat y,\hat t;\lb)=\sum_{0\leq i\leq
p-1}a_i\Phi^{(1)}(\om^i\lb)+\sum_{0\leq i\leq
p-1}b_i\Phi^{(2)}(\om^i\lb)
\ee
flowing according to the equations (0.20); in (6.3)
\begin{eqnarray*}
\Phi^{(1)}(\hat x,\hat y,\hat
t;\lb)&:=&\Psi_1(x,y;\lb)e^{-\sum_1^{\iy}y_{ip}\lb^{ip}}\\
&=&e^{\sum_{i\not| p}x_i\lb^i}e^{\sum_{i=1}^{\iy}t_{ip}\lb^{ip}}\left(
z^n\frac{\tau_n(\hat x-[\lb^{-1}],\hat y,\hat
t-[\lb^{-1}])}{\tau_n(\hat x,\hat y,\hat t)}\right)_{n\in\BZ}\\
&=&\frac{\BX_1(\lb)\tau}{\tau}
\end{eqnarray*}
\medbreak
\begin{eqnarray*}
\Phi^{(2)}(\hat x,\hat y,\hat
t;\lb)&:=&\Psi_2(x,y;\lb^{-1})e^{-\sum_1^{\iy}y_{ip}\lb^{ip}}\\
&=&e^{\sum_{i\not| p}y_i\lb^i}\left(
z^{-n}\frac{\tau_{n+1}(\hat x,\hat y-[\lb^{-1}],\hat
t+[\lb^{-1}])}{\tau_n(\hat x,\hat y,\hat t)}\right)_{n\in\BZ}\\
&=&\frac{e^{\sum_1^{\iy}t_{ip}\lb^{ip}}\BX_2(\lb)\tau}{\tau}.
\end{eqnarray*}

\proof The proof proceeds as in section 2; for instance, (2.2) gets
replaced by
$$
\frac{\pl(L_1^p-L_2^p)}{\pl x_n}=0,\quad\frac{\pl(L_1^p-L_2^p)}{\pl
y_n}=0
$$
and (2.3) by
$$
\left(\frac{\pl}{\pl
x_{np}}+\frac{\pl}{\pl
y_{np}}\right)L_i=[(L_1^{np})_++(L_2^{np})_-,L_i]=
[(L_i^{np})_++(L^{np})_-,L_i]=0.
$$
That $\Phi$ satisfies the differential equations (0.20) proceeds along
the same lines as (2.9) and (2.10).  \qed

\medbreak

\underline{\sl Proof of Theorem 0.4}: Lemma 2 implies Theorem 0.4,
except for the statement about the Darboux transform. For that, one
needs to factor $L$ according to the recipe of Lemma 1, where we set
$\Phi=\Phi^{(1)}$, in the precise notation of Lemma 1. Thus the
Darboux transformation corresponds to bringing the factor most to the
right all the way to the left.\qed

\end{document}